\documentclass[article]{aastex63}

\definecolor{DarkerGreen}{rgb}{0.0,0.7,0.1}
\newcommand{\revtwo}[1]{{\bf\color{black} #1}}

\newcommand{\ebtel}{{\tt EBTEL}}
\newcommand{\hydrad}{{\tt HYDRAD}}
\newcommand{\ebtell}{{\tt EBTEL2}}

\newcommand{\ebtelll}{{\tt EBTEL3}}
\newcommand{\ebtelke}{{\ebtell}+\tt KE}

\usepackage{appendix,amsmath,ulem}
\received{}
\revised{}
\accepted{}
\submitjournal{ApJ}

\shorttitle{Flows in EBTEL}
\shortauthors{}
\begin{document}
\title{Flows in Enthalpy Based Thermal Evolution of Loops}
\correspondingauthor{Abhishek Rajhans}
\email{abhishek@iucaa.in}
\author[0000-0001-5992-7060]{Abhishek Rajhans}
\author[0000-0003-1689-6254]{Durgesh Tripathi}

\affiliation{Inter-University Centre for Astronomy and Astrophysics, Post Bag - 4, Ganeshkhind, Pune 411007, India}

\author[0000-0002-3300-6041]{Stephen J. Bradshaw}
\affiliation{Rice University, Department of Physics and Astronomy, 6100 Main Street Houston, TX 77004, USA}

\author[0000-0002-3869-7996]{Vinay L. Kashyap}
\affiliation{Center for Astrophysics $|$ Harvard \& Smithsonian, 60 Garden St.\ Cambridge MA 02138, USA }

\author[0000-0003-2255-0305]{James A. Klimchuk}
\affiliation{NASA Goddard Space Flight Center, Solar Physics Laboratory, Code 671, Greenbelt, MD 20771, USA}

\begin{abstract}
Plasma filled loop structures are common in the solar corona.  Because detailed modeling of the dynamical evolution of these structures is computationally costly, an efficient method for computing approximate but quick physics-based solutions is to rely on space integrated 0D simulations. The enthalpy based thermal evolution of loops (\ebtel) framework is a commonly used method to study the exchange of mass and energy between the corona and transition region. {\ebtel} solves for density, temperature, and pressure, averaged over the coronal part of loop, velocity at coronal base, and the instantaneous differential emission measure distribution in the transition region. The current single-fluid version of the code, {\ebtell}, assumes that at all stages the flows are subsonic.
However, sometimes the solutions show the presence of supersonic flows during the impulsive phase of heat input. It is thus necessary to account for this effect. Here, we upgrade {\ebtell} to {\ebtelll} by including the kinetic energy term in the Navier-Stokes equation.
We compare the solutions from {\ebtelll} with those obtained using {\ebtell} as well as
the state-of-the-art field aligned hydrodynamics code {\hydrad}. We find that the match in pressure between {\ebtelll} and {\hydrad} is better than that between {\ebtell} and {\hydrad}. Additionally, the velocities predicted by {\ebtelll} are in close agreement with those obtained with {\hydrad} when the flows are subsonic. However, {\ebtelll} solutions deviate substantially from \hydrad's when the latter predicts supersonic flows. Using the mismatches in the solution, we propose a criterion to determine the conditions under which {\ebtel} can be used to study flows in the system.
\end{abstract}


\section{Introduction} \label{sec:intro}

The Solar corona has a myriad of loop like structures, which are magnetic flux tubes filled with plasma. Field aligned hydrodynamic simulations serve as an efficient tool for studying the response of plasma in coronal loops to a generic time dependent heating event, \citep[see eg.,][for a review]{Klimchuk2006, rspajim, reale}. However in situations where one needs to study the effects of variation of parameters like loop length, energy budget and profile of heating function, a large number of runs are required. Additionally more realistic scenarios involve multi-stranded loops as opposed to simple monolithic ones. Performing field aligned simulations for such complicated but realistic scenarios is 
computationally expensive. To overcome such issues, 0D codes have been developed. The idea is that in the coronal part, physical parameters do not vary much along the loop. Hence the values of average density, temperature and pressure are characteristic for the coronal part of the loop \citep[see][for a detailed discussion on various 0D models]{cargill0dmodels}. In this work we restrict ourselves to the commonly used 0D code: Enthalpy Based Thermal Evolution of Loops \citep[{\ebtel}; see][for details]{Klimchuk2008, cargill, Barnes_2016}.

In a nutshell, {\ebtel} studies the mass and energy exchange between the corona and the transition region. It calculates average temperature, density, pressure, and velocity at the base of the coronal loops. It also computes the instantaneous differential emission measure in the transition region. The  coronal base of a loop is defined as the position at which thermal conduction changes from cooling term to heating term. Mathematically, this is the point where the second spatial derivative of temperature changes its sign. From field aligned simulations, it has been shown that such change in the sign, occurs at a position along the loop where the temperature is approximately 0.6 times the temperature at loop apex \citep[][]{Klimchuk2008}. \cite{cargill} included additional physics such as gravitational stratification and corrections in the radiative cooling phase, 
for studying the evolution of single fluid plasma in loops, in the 0D description developed in \cite{Klimchuk2008}. The version of single-fluid {\ebtel} used in \cite{cargill} will be referred to as {\ebtell} throughout. The results obtained using {\ebtell} show remarkable agreement with those obtained from the field aligned simulations performed with state-of-the-art HYDrodynamics and RADiation ({\hydrad}) \citep[][]{bradshaw2003, bradshaw2006} code and differ at most by 15{--}20\% \citep[][]{cargill}. 

One of the crucial assumptions made in {\ebtel} and {\ebtell} is that at all stages  of the loop evolution, the flows are subsonic. Therefore, the kinetic energy term is neglected from the Navier-Stokes equation. The assumption is consistent with results from field aligned simulations performed using wide range of loop length and heating function. However in cases where {\hydrad} predicts subsonic flows, {\ebtell} may compute Mach numbers close to, or even exceeding unity. This is a consequence of neglecting the kinetic energy, which  becomes relevant during the impulsive phase of evolution of loop in these cases. Mach numbers exceeding unity in reality would lead to shocks and disrupt flows by converting kinetic energy into heat. In such situations, a simplified 0D description becomes inadequate. Therefore, it is necessary to include the kinetic energy term in the Navier-Stokes equation to be able to study the dynamics of the loop. This is the main aim of this paper.

It is important to point out that the Mach numbers generated by field aligned codes can also reach values close to or exceeding unity. Note that while field aligned codes such as {\hydrad} have the potential to tackle shocks, 
that is not the case for 0D codes such as {\ebtel}. Nevertheless, despite the simplified nature of 0D calculations, the computed temperature, density, and pressure are in reasonable agreement with field aligned simulations performed using {\hydrad}.

The rest of the paper is structured as follows. In \S~\ref{sec:ebtelframework} we describe the 0D description of coronal loops, first neglecting kinetic energy in \S~\ref{sec:notincludeke}, and then including it in \S~\ref{sec:includeke}.
We discuss the results for a particular exemplar case. The 
0D code based on modified equations derived in \S~\ref{sec:includeke} is referred to as {\ebtelke}. We compare these results with those obtained using {\ebtell} as well as using field aligned solution from {\hydrad}. In \S~\ref{sec:densratio}, we discuss the validity of approximations used in \S~\ref{sec:includeke}, describe their shortcomings, improve upon these approximations, and demonstrate their validity. All the changes discussed in \S~\ref{sec:includeke} and \ref{sec:densratio} are incorporated in a newer version of the code, {\ebtelll}. In \S~\ref{subsec:case2}-\ref{subsec:case7} we show the results obtained from {\ebtelll} for various cases, covering wide range of loop length, heating function, and Mach numbers at coronal base. These have been compared with results obtained from {\ebtell} and {\hydrad}. We also investigate the parameter space where the simulated 0D Mach numbers are unreliable and develop a heuristic way to identify such instances in \S~\ref{subsec:predict}. Finally, we summarize our work in \S~\ref{sec:disc}.

\section{EBTEL Framework}\label{sec:ebtelframework}

The  
0D description of coronal loops requires integrating the field aligned hydrodynamical equations over the corona and the transition region. We first describe the field aligned equations in brief.

The Navier Stokes equation for inviscid fluid for high temperature plasma filling a magnetic flux tube is, 
\begin{equation}\label{eq:energy1dkin}
\frac{\partial}{\partial t}\left(\frac{1}{\gamma-1}P+\frac{1}{2}n\mu v^{2}\right) = -\frac{\partial}{\partial s}\left(\left[\frac{\gamma}{\gamma-1}P+\frac{1}{2}n\mu v^{2}\right]v+F\right)+Q-n^{2}\Lambda(T)+n\mu g_{||}v \,,
\end{equation}
where $t$ denotes time, $s$ denotes field aligned loop coordinate, $P$ is pressure, $\gamma$ is the polytropic index, $v$ is velocity, $\mu$ is the effective ion mass, $n$ is the electron number density, $Q$ is the heating rate per unit volume,  $F$ is the heat flux, 
$\Lambda(T)$ is the optically thin power loss function, $T$ is the temperature,  
and $g_{||}$ is the component of acceleration due to gravity along the direction of the magnetic field \citep[see e.g.][]{Klimchuk2008}.

The equation of motion is,
\begin{equation}\label{eq:eqofmotion}
\frac{\partial v}{\partial t}+v\frac{\partial v}{\partial s}=-\frac{1}{n\mu}\frac{\partial P}{\partial s}+g_{||} \,,
\end{equation}

\noindent and the equation governing conservation of mass is,
\begin{equation}\label{eq:mass1d}
\frac{\partial n}{\partial t} + \frac{\partial J}{\partial s}  = 0   
\end{equation}
\noindent where $J$ is the electron flux.  

These equations are closed by an equation of state which in this case is the ideal gas law 
$P = 2nk_{B}T$, where $k_{B}$ is the Boltzmann constant. The factor of 2 is because the plasma is assumed to be fully ionized hydrogen plasma. While $P$ refers to total pressure of ions and electrons, $n$ is number density of electrons.

\subsection{0D equations for subsonic flows}\label{sec:notincludeke}

 \cite{Klimchuk2008} derived the 0D equations for subsonic flows by neglecting the kinetic energy term ($\frac{1}{2}n\mu v^{2} << P$)
in a symmetric loop with a uniform cross section.  Due
to symmetry, the vector quantities, {\it viz.} heat flux, electron number density flux, and velocity vanish at the loop apex. 

On integrating equation~\ref{eq:energy1dkin} under the assumption of subsonic flows, from the coronal base of the loop ($s=0$) to the loop apex ($s=L$), we find 
\begin{equation}\label{eq:0denergycorona_newsubsonic}
\frac{1}{\gamma-1}L\frac{d\bar{P}}{dt}=\frac{\gamma}{\gamma-1}P_{0}v_{0}+F_{0}+\bar{Q}L-\bar{n}^{2}\Lambda(\bar{T})L \,,
\end{equation}
 
\noindent where $\bar{P}$, $\bar{n}$, $\bar{T}$ and $\bar{Q}$ denote the pressure, electron number density, temperature, and heating rate per unit volume, averaged over the coronal part of the loop respectively. $P_{0}$ and $v_{0}$ are the pressure and velocity respectively at the coronal base. $F_{0}$ is the heat flux across $s=0$. 

The heat flux is due to particles in the thermal pool. This is a combination of Spitzer flux accounting for Coulomb 
collisions, and saturation flux that corrects for over-prediction of 
the heat flux by classical expression \citep[see][]{Klimchuk2008}. 
However, a significant fraction of energy can also be carried away by non-thermal 
electrons. The effect of the latter can be readily included as discussed in \cite{Klimchuk2008}. For this purpose, flux of non-thermal electrons across coronal base ($J_{nt0}$) and average energy per non-thermal electron $(E_{nt})$ need to be given as additional inputs.  Under such 
a scenario, $F_{0}$ is the sum of the thermal conduction flux ($F_{t0}$) and the energy flux  carried by non-thermal electrons ($E_{nt}J_{nt0}$), across the coronal base. Similarly $J_{0}$ becomes the sum of flux of electrons in the thermal pool ($n_{0}v_{0}$), and non-thermal electrons ($J_{nt0}$), across the coronal base.

On integrating equation~\ref{eq:energy1dkin} for subsonic flows, from the base of the part of loop in transition region ($s=-l$) to the coronal base of the loop ($s=0$),  we find 
\begin{equation}\label{eq:0denergytr_newsubsonic}
\frac{1}{\gamma-1}l\frac{d \bar{P}_{tr}}{dt}\approx 0 =-\frac{\gamma}{\gamma-1}P_{0}v_{0}-F_{0}+l\bar{Q}-c_{1}\bar{n}^{2}\Lambda(\bar{T})L \,,
\end{equation}
where $\bar{P}_{tr}$ is the average pressure in the transition   
region, and following \citet{Klimchuk2008} we ignore $l\frac{d \bar{P}_{tr}}{dt}$ and $l\bar{Q}$ because of small length of the transition region ($l$).

The quantity $c_{1}\bar{n}^{2}\Lambda(\bar{T})L$ is the total radiation loss from transition region, where $\bar{n}$ is the average electron density in corona (not the transition region). This is $c_{1}$ times the total radiation loss from the corona. Based on results from field aligned simulations, it was assumed to be constant in time and equal to 4 in \cite{Klimchuk2008}. However \cite{cargill} dynamically computed it at each time step by including gravitational stratification of corona and corrections in radiative cooling phase, when the loops become over dense as compared to equilibrium value. Both {\ebtell} and enhancement developed in this work ({\ebtelll}), compute $c_{1}$ dynamically at each time step. 

Adding equations \ref{eq:0denergycorona_newsubsonic} and \ref{eq:0denergytr_newsubsonic}, we get
\begin{equation}\label{eq:newfpressure0dsubsonic}
\frac{1}{\gamma-1}L\frac{d\bar{P}}{dt}= \bar{Q}L -(1+c_{1})\bar{n}^{2}\Lambda(\bar{T})L \,.
\end{equation}
On integrating equation~\ref{eq:mass1d} from 
the base of the loop in the corona 
($s$ = $0$) to 
the top of the loop  
($s = L$), we get
\begin{equation}\label{eq:zerodmass}
L\frac{d\bar{n}}{dt} = J_{0} = n_{0}v_{0}=  \frac{P_{0}}{2k_{B}T_{0}}v_{0} \,,
\end{equation}
where $n_{0}$ is the  electron number density at coronal base, $J_{0}$ is the flux of electrons across the coronal base ($s=0$). If there is an additional flux of non-thermal  electrons, $J_{0} = \frac{P_{0}}{2k_{B}T_{0}}v_{0} +J_{nt0}$. In either case for expressing $P_{0}$ in terms of $\bar{P}$, it is assumed that the loop is in hydrostatic equilibrium and is isothermal, with temperature equal to the average coronal temperature, i.e. $\left[\frac{P_{0}}{\bar{P}}\right] = \left[\frac{P_{0}}{\bar{P}}\right]_{hse}$. The subscript $hse$ denotes hydrostatic equilibrium and at uniform temperature. The temperature at loop top ($T_{a}$) is related to $\bar{T}$ by a constant $c_{2} = 0.9$, such that $\frac{\bar{T}}{T_{a}} = c_{2}$.  The temperature at coronal base ($T_{0}$) is related to temperature at loop top ($T_{a}$) by another constant $c_{3} = 0.6$, such that $\frac{T_{0}}{T_{a}} = c_{3}$. 

Finally, the system described by equations.~\ref{eq:0denergytr_newsubsonic}, \ref{eq:newfpressure0dsubsonic}, and \ref{eq:zerodmass} is closed by the ideal gas law,
\begin{equation}\label{eq:eos}
\bar{P} = 2\bar{n}k_{B}\bar{T}     
\end{equation}
  
\subsection{0D equations without the assumption of subsonic flows}\label{sec:includeke}
We now derive the 0D equations that avoid the assumption of subsonic flows in coronal loops. Using 
equations~\ref{eq:eqofmotion} and \ref{eq:mass1d}, 
we 
write the time derivative of kinetic energy,

\begin{equation}\label{eq:kinetic_td}
\frac{\partial}{\partial t}\left(\frac{1}{2}n\mu v^{2}\right) = -\frac{1}{2}\mu v^{2}\frac{\partial }{\partial s}(J-nv)- \frac{\partial}{\partial s}\left(\frac{1}{2}n \mu v^{3}\right) -v\frac{\partial P}{\partial s}+n\mu vg_{||} \,.    
\end{equation}

Plugging equation~\ref{eq:kinetic_td} in \ref{eq:energy1dkin}, we obtain

\begin{equation}\label{eq:final1denergyeq1}
\frac{1}{\gamma-1}\frac{\partial P}{\partial t}=\frac{1}{2}\mu v^{2}\frac{\partial }{\partial s}(J-nv) + v\frac{\partial P}{\partial s}-\frac{\partial}{\partial s}\left(\frac{\gamma}{\gamma-1}Pv+F\right)+Q-n^{2}\Lambda(T) \,.
\end{equation}

As discussed in \S~\ref{sec:notincludeke}, we can include the effect of non-thermal electrons by replacing $J$ with $nv+J_{nt}$, and $F$ with $F_{t} + E_{nt}J_{nt}$. This gives,

\begin{equation}\label{eq:final1denergyeq2}
\frac{1}{\gamma-1}\frac{\partial P}{\partial t}=\left(\frac{1}{2}\mu v^{2}-E_{nt}\right)\frac{\partial }{\partial s}(J_{nt}) + v\frac{\partial P}{\partial s}-\frac{\partial}{\partial s}\left(\frac{\gamma}{\gamma-1}Pv+F_{t}\right)+Q-n^{2}\Lambda(T) \,.
\end{equation}

Characteristic velocity of $10^{6-7}$~cm~s$^{-1}$ in corona gives kinetic energy of an ion ($\sim \mu v^{2}$) $\approx 10^{-12} - 10^{-10}$~ergs. This is similar to average thermal energy per electron ($\sim k_{B}T$) $\approx 10^{-10}$~ergs for $T$=1~MK. For the non-thermal electrons to escape the thermal pool, their average energy should be much larger than the average thermal energy per 
electron ($\frac{1}{2}\mu v^{2}$), and hence the above equation~\ref{eq:final1denergyeq2} can be simplified, 
as 
\begin{equation}\label{eq:final1denergyeq3}
\frac{1}{\gamma-1}\frac{\partial P}{\partial t}= v\frac{\partial P}{\partial s}-\frac{\partial}{\partial s}\left(\frac{\gamma}{\gamma-1}Pv+F\right)+Q-n^{2}\Lambda(T) \,.
\end{equation}

Integrating equation~\ref{eq:final1denergyeq3} from the coronal base of the loop ($s=0$) to the loop apex ($s=L$) we find,
\begin{equation}\label{eq:0denergycorona_new}
\frac{1}{\gamma-1}L\frac{d\bar{P}}{dt}=\int_{0}^{L}v\frac{\partial P}{\partial s}ds+\frac{\gamma}{\gamma-1}P_{0}v_{0}+F_{0}+\bar{Q}L-\bar{n}^{2}\Lambda(\bar{T})L \,.
\end{equation}
 
Similarly on integrating  equation~\ref{eq:final1denergyeq3} from base of part of the loop in the transition region $(s=-l)$ to the coronal base of the loop $(s=0)$, we have

\begin{equation}\label{eq:0denergytr_new}
\frac{1}{\gamma-1}l\frac{d \bar{P}_{tr}}{dt}\approx 0 =\int_{-l}^{0}v\frac{\partial P}{\partial s}ds-\frac{\gamma}{\gamma-1}P_{0}v_{0}-F_{0}-c_{1}\bar{n}^{2}\Lambda(\bar{T})L \,.
\end{equation}

To solve 
the integral on 
the 
right hand side of equation~\ref{eq:0denergytr_new}, we first multiply the equation of motion 
(equation~\ref{eq:eqofmotion}) 
with $-n\mu v$ throughout and then integrate 
from $s=0$ to $s=-l$. This gives us

\begin{equation}\label{eq:integal_tr}
\int_{-l}^{0}v\frac{\partial P}{\partial s}ds=-\int_{-l}^{0}\left(n\mu v\frac{\partial v}{\partial t}+n\mu v^{2}\frac{\partial v}{\partial s}-n\mu  vg_{||}\right)ds \approx -\frac{1}{2}n_{0}\mu v_{0}^{3} \,.
\end{equation}

\noindent where we have assumed that the transition region is in a steady state, i.e. $\frac{\partial v}{\partial t} = 0$ and $nv = n_{0}v_{0}$. 
Using
estimates of 
$g \approx 10^4$~cm~s$^{-2}$, $l \approx 10^7$~cm, and $v_{0} \approx 10^{6-7}$~cm~s$^{-1}$ 
\citep[][]{Klimchuk2008}, we find that $n_{0}\mu v_{0}g_{||}l$ is negligible in comparison to $\frac{1}{2}n_{0}\mu v_{0}^{3}$.

Using equations~\ref{eq:integal_tr} in \ref{eq:0denergytr_new} we obtain (see also \cite{priest2014magnetohydrodynamics})

\begin{equation}\label{eq:0denergytr_newfinal}
0 = -\frac{1}{2}n_{0}\mu v_{0}^{3} -\frac{\gamma}{\gamma-1}P_{0}v_{0}-F_{0}-c_{1}\bar{n}^{2}\Lambda(\bar{T})L \,,
\end{equation}

\noindent which 
refers to the statement of energy balance in transition region. The balance of 
the 
first three terms 
i.e., 
fluxes of energy (kinetic, enthalpy, and heat) passing between the transition region and corona, is the energy lost by radiation. It has been assumed that any mass and energy flow through the bottom of transition region is negligible. Work  done by or against gravity, and direct heating in the thin transition region have also been neglected. However it should be noted that in the presence of 
a 
large variation in cross sectional area, 
the 
transition region can thicken substantially \citep[][]{cargill2021}. 
The kinetic energy term in equation~\ref{eq:0denergytr_newfinal} is the only difference from the earlier versions of EBTEL.  
Note that equation~\ref{eq:0denergytr_newfinal} 
is cubic in $v_0$, 
and 
can be solved analytically (see appendix~\ref{app:cubic}). 

Adding equation~\ref{eq:0denergytr_newfinal}  and equation~\ref{eq:0denergycorona_new} we get     
 
\begin{equation}\label{eq:newpressure0d}
\frac{1}{\gamma-1}L\frac{d\bar{P}}{dt}=-\frac{1}{2}n_{0}\mu v_{0}^{3} + \int_{0}^{L}v\frac{\partial P}{\partial s}ds + \bar{Q}L -(1+c_{1})\bar{n}^{2}\Lambda(\bar{T})L \,.
\end{equation}
 
\noindent Field aligned hydrodynamic simulations performed using {\hydrad} for benchmarking  {\ebtelll} show that 
\begin{equation}\label{eq:secondapprox}
\int_{0}^{L}v\frac{\partial P}{\partial s}ds \approx -\frac{1}{2}n_{0}\mu v_{0}^{3}    
\end{equation}
holds well during most of the times. It should be noted that this integral under steady state assumption in corona, should have been $\int_{0}^{L}v\frac{\partial P}{\partial s}ds \approx -\int_{0}^{L}\left(n\mu v\frac{\partial v}{\partial t}+n\mu v^{2}\frac{\partial v}{\partial s}-n\mu  vg_{||}\right)ds \approx \frac{1}{2}n_{0}\mu v_{0}^{3}$, if the  third term was ignorable in the corona as it is in the transition region (see equation~\ref{eq:integal_tr}). This gives us the same magnitude as the expression we use, however the sign is opposite. Plugging this value in equation~\ref{eq:newpressure0d}, we obtain

\begin{equation}\label{eq:newfpressure0d}
\frac{1}{\gamma-1}L\frac{d\bar{P}}{dt}=-n_{0}\mu v_{0}^{3}  + \bar{Q}L -(1+c_{1})\bar{n}^{2}\Lambda(\bar{T})L \,.
\end{equation}
The description of system is completed by equations~\ref{eq:zerodmass} and \ref{eq:eos}.

\begin{figure}[ht]
\centering 
\includegraphics[width=\linewidth]{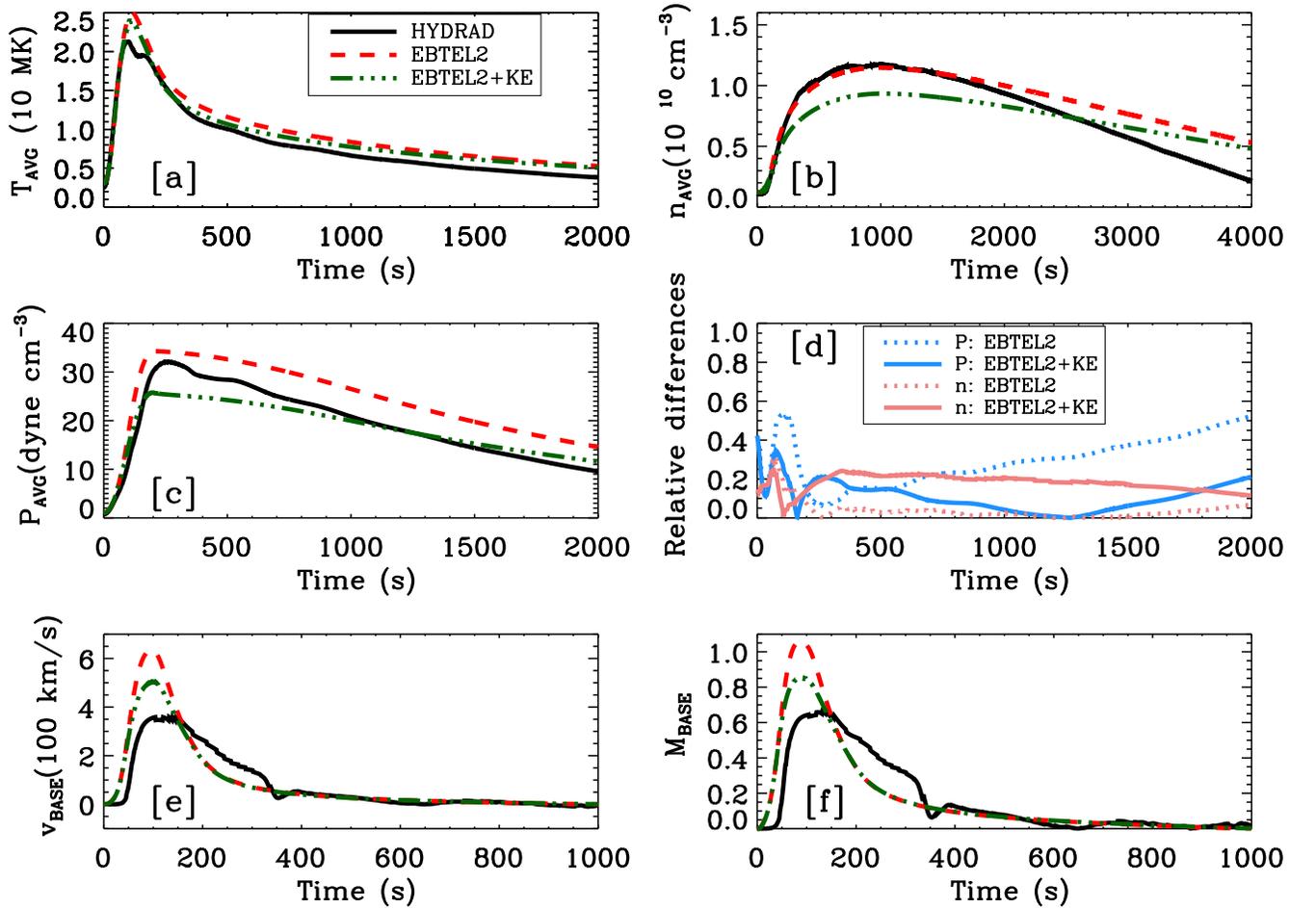} 
\caption{Simulation results for the exemplar case. The curves demonstrate the  time evolution of  $\bar{T}$ (panel [a]), $\bar{n}$ (panel [b]), and $\bar{P}$ (panel [c]), the discrepancy in $\bar{n}$ (red curves) and $\bar{P}$ (blue curves) between \hydrad\ and those calculated from {\ebtell} (dotted) and {\ebtelke} (solid) (panel [d]), velocity at coronal base of loop ($v_{0}$) (panel [e]), and corresponding Mach number ($M_{0}$) (panel [f]).} \label{fig:figcase0} 
\end{figure}

In order to test the results obtained by equations~ \ref{eq:zerodmass}, \ref{eq:eos}, \ref{eq:0denergytr_newfinal}, and \ref{eq:newfpressure0d}, we consider an exemplar case which generates supersonic flows in {\ebtell} and subsonic flows in {\hydrad}. This serves as a useful test for simulations performed using the system of modified equations~\ref{eq:zerodmass}, \ref{eq:eos}, \ref{eq:0denergytr_newfinal}, and \ref{eq:newfpressure0d},  which are collectively labeled as {\ebtelke}. We simulate the evolution of a loop of half length 65~Mm, with background heating adjusted such that the initial density and temperature are $11.05\times10^{8}$~cm$^{-3}$ and 2.51~MK, respectively. The system is subjected to a symmetric triangular heating profile lasting for 200~s, with the maximum heating rate being 0.5 ergs~cm$^{-3}$~s$^{-1}$ at $t=100$ s. This corresponds to energy deposition of 3.25 $\times 10^{11}$ ergs~cm$^{-2}$ in 200 s.

The results of this test case are shown in Figure~\ref{fig:figcase0},
with simulations from {\hydrad} (solid black curves), {\ebtell} (dashed red curves), and {\ebtelke}  (dash-triple-dotted green curves). Following \citet{cargill} and \citet{Klimchuk2008}, we have identified the coronal part of the loop in {\hydrad} as the portion where the temperature is greater than or equal to 0.6 times the temperature at loop top. Pressure, density, and temperature have been averaged over this region to obtain coronal averages. The lowest point of this region is taken as the coronal base, where velocity and Mach number have been evaluated. The apex quantities have been obtained by {\hydrad}, averaging over the top 20\% of the length of the loop.  

Figure~\ref{fig:figcase0} shows the time evolution of average temperature (panel [a]),  the average electron number density (panel [b]), the average pressure (panel [c]).  We also plot in panel [d] the absolute relative differences of $\bar{P}$ and $\bar{n}$ as computed from {\ebtell} or {\ebtelke} relative to HYDRAD, which are defined as 
\begin{equation}
\frac{\Delta{\bar{P}}}{\bar{P}} = \left|\frac{\bar{P}_{\hydrad}-\bar{P}_{\ebtel}}{\bar{P}_{\hydrad}}\right|   
\hspace{0.3cm} \&  \hspace{0.3cm}
\frac{\Delta{\bar{n}}}{\bar{n}}  = \left|\frac{\bar{n}_{\hydrad}-\bar{n}_{\ebtel}}{\bar{n}_{\hydrad}}\right| \,.
\end{equation}
\noindent In panels [e] and [f], we show the time  evolution of velocity and Mach number at coronal base ($v_{0}$ and $M_{0}$) respectively.

We find that the temperatures obtained using {\ebtell} and {\ebtelke} are nearly identical (panel [a]). The electron number densities obtained with {\ebtelke} are, however, consistently lower than those obtained with {\ebtell} (panel [b]). 
However, the relative error remains less than 25 \% (see panel [d]). The average pressure plotted in panel [c] shows that the agreement between {\hydrad} and  {\ebtelke} is better, except during a small time window close to where pressure peaks. A better agreement for velocity is also seen in {\hydrad} and {\ebtelke} (panel [e]) and Mach number (panel [f]) at the base. While {\hydrad} produces subsonic flows at base (panel [f]), {\ebtell} produces Mach numbers exceeding 1. If these are trusted, it should lead to shocks, which cannot be tackled by 0D simulations. The {\ebtelke} brings down Mach numbers below unity. However, we note that improvements in Mach numbers are unsatisfactory, given the small improvement in velocity and Mach numbers at the cost of deterioration in electron number density. Therefore, further improvements are needed.

\subsection{Assessment and improvements of the approximations}\label{sec:densratio} 

While seeking 
solutions to the equations~\ref{eq:zerodmass}, \ref{eq:eos}, \ref{eq:0denergytr_newfinal}, and \ref{eq:newfpressure0d}, we have made two approximations, which we now discuss in some details. The 
first approximation (also 
present in  {\ebtell}) is related to the ratio of pressure and electron number density at the coronal base with their average values. {\ebtell} computes the ratio of the pressure at the coronal base and average pressure by assuming the system to be in hydrostatic equilibrium i.e. $P(s) = P_{0}\exp(-s/L_{H}(\bar{T}))$. 
The variation of temperature along the loop is neglected and scale height ($L_{H}$) is computed using a  temperature equal to the coronal average value. \cite{cargill} discuss that this is equivalent to $[\frac{P_{0}}{\bar{P}}]_{hse} = \exp(2L\sin({\frac{\pi}{5}})/\pi L_{H}(\bar{T}))$ for a semicircular loop, where the subscript hse indicates that the loop is in hydrostatic equilibrium at uniform temperature ($\bar{T}$). Using the constants $\frac{\bar{T}}{T_{a}} = c_{2} = 0.9$ and $\frac{T_{0}}{T_{a}} = c_{3} = 0.6$, the temperature at the coronal base ($T_{0}$) is 0.67 times the average temperature of loop ($\bar{T}$). Using these along with $P_{0} = 2n_{0}k_{B}T_{0}$ and $\bar{P} = 2\bar{n}k_{B}\bar{T}$, the ratio of electron number density at the coronal base to its coronal average ,
\begin{equation}
   \left[\frac{n_{0}}{\bar{n}}\right]_{hse} = \frac{3}{2}\exp(2L\sin({\frac{\pi}{5}})/\pi L_{H}(\bar{T})) \,.
\end{equation}

To assess this approximations, we define the following two coefficients and compute these using {\hydrad},
\begin{equation}
c_{4} = \left[\frac{P_{0}}{\bar{P}}\right] \left[\frac{P_{0}}{\bar{P}}\right]_{hse}^{-1}  \hspace{0.3cm} \& \hspace{0.3cm} c_{5} = \left[\frac{n_{0}}{\bar{n}}\right] \left[\frac{n_{0}}{\bar{n}}\right]_{hse}^{-1} 
\label{eq:c4c5}
\end{equation}

\begin{figure}[ht]
\centering 
\includegraphics[width=0.5\linewidth]{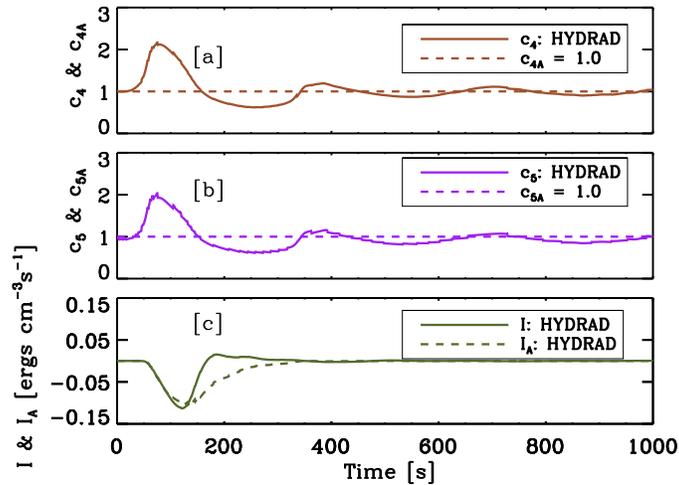} 
\caption{Comparing the effects of different approximations within \hydrad\ for the exemplar case (see Figure~\ref{fig:figcase1}).  {\bf [a]} The {\sl top panel} (with brown curves) shows the evolution of the pressure coefficient $c_{4}$ (solid curve; equation~\ref{eq:c4c5}) and the approximation used in \ebtell, $c_{4A}=1$ (dashed line).  {\bf [b]} The {\sl middle pane} (purple curves) shows the evolution of the electron number density coefficient $c_{5}$ (solid curve; equation~\ref{eq:c4c5}) and the approximation used in \ebtell, $c_{5A}=1$ (dashed line). {\bf [c]} The {\sl bottom panel} (olive curves) shows the evolution of energy flux $I$ and our approximation $I_{A}$ (solid and dashed curves respectively; equation~\ref{eq:secondapproxarray}), with both $I$ and $I_{A}$ are computed using \hydrad.  Note that the approximations to the pressure and electron number density coefficients $c_{4A}$ and $c_{5A}$ are discrepant by factors of up to 2, whereas the energy flux approximation $I_{A}$ is adequate.}
\label{fig:figcase01} 
\end{figure}

We show the evolution of $c_{4}$ and $c_{5}$ as the solid curves in panels [a] and [b] of Figure~\ref{fig:figcase01} respectively.  In order to relate the pressure and electron number density at the base of the corona and their averages across the loop, \ebtell\ assumes a hydrostatic profile, with $c_{4}$ and $c_{5}$ approximated as constants, $c_{4A}=1$ and $c_{5A}=1$ respectively; these are shown as dashed lines in Figure~\ref{fig:figcase01}.  These approximation are clearly discrepant by up to factors of 2 during the impulsive phase.  Hence, we need to develop a better approximation to relate the base pressure and electron number density to the quantities computed within the \ebtel\ framework, $\bar{P}, \bar{n}, \bar{T} $ and $v_{0}$

We attribute the variation of $c_{4} $ and $c_{5}$ from their constant values of 1, mainly to two aspects: 1) since the plasma is accelerated into the loop, it requires larger pressure gradients than hydrostatic values, in lower parts of loop, 2) since the plasma velocity at apex of the loop is zero, the flow must be decelerated, and therefore the pressure gradient in the upper parts of the loop should be smaller than hydrostatic or perhaps even in the opposite direction. From the plots, it appears that for the first 150~s, the first cause dominates, while the second dominates during the later phase till  about 350~s.

\noindent To ascertain the validity of our second approximation (equation~\ref{eq:secondapprox}), we compute following two quantities,
\begin{equation}\label{eq:secondapproxarray}
\begin{aligned}
I =  \int_{0}^{L}v\frac{\partial P}{\partial s}ds  -\frac{1}{2}n_{0}\mu v_{0}^{3} ~~~~~~{\rm (without~approximation)} \\
I_{A} =  -n_{0}\mu v_{0}^{3} ~~~~~~{\rm (with~approximation)} 
\end{aligned}
\end{equation}
\noindent using {\hydrad} and plot them in panel [c] of figure~\ref{fig:figcase01}. The reasonable match between the two curves suggests that this approximation is satisfactory, albeit being relatively poor between 150-350 seconds. This could be attributed to the following effects. In the impulsive phase (upto~150 s) in order to fill plasma into the loop, there should be acceleration. Hence dominating contribution to $I$ should be from portions of loop where force due to pressure gradients $(-\frac{\partial  P}{\partial s})$ are  in the same direction as the velocity of plasma ($v$). This means pressure gradients are opposite to plasma velocity at locations from where  larger contribution to $I$ comes from. Since our approximation $I_{A}$ predicts negative values, it matches well with $I$ in this duration. However, between 150-350 seconds, the effect of pressure gradients with directions opposite to hydrostatic pressure gradients (for decelerating plasma) becomes more important (see panel [a] of figure~\ref{fig:figcase01}). Since plasma is being decelerated, larger contribution to $I$ should come from portions of the loop where pressure force $(-\frac{\partial P}{\partial s})$ is opposite to velocity $(v)$. Clearly for such regions pressure gradient is along velocity, but since our approximation $I_{A}$ still gives negative values, it introduces errors.

In order to improve on the first approximation, we first develop 
insight on the form of pressure at base of corona in terms of quantities computed within 
the framework of 0D simulations under assumption of steady flow and uniform temperature ($\bar{T}$) along the loop. While these conditions are certainly not met in the impulsive phase where kinetic energy dominates, the resulting expressions can be generalized. 

Assuming a steady flow in equation~\ref{eq:eqofmotion} and combining that with equation~\ref{eq:eos}, we have

\begin{equation}\label{eq:steadyflow1}
\frac{1}{P}\frac{\partial P}{\partial s} =  \frac{\mu}{2k_{B}\bar{T}}\left(g_{||}-\frac{1}{2}\frac{\partial v^{2}}{\partial s}\right) 
\end{equation}

Under the assumption that acceleration due to gravity and any variation in temperature can be neglected, we integrate equation~\ref{eq:steadyflow1} to obtain

\begin{equation}\label{eq:steadyflow2}
\frac{P(s)}{P_{0}} = \exp\left[\frac{\mu}{2k_{B}\bar{T}}\left(sg_{||}-\frac{1}{2}(v^{2}-v_{0}^{2})\right)\right] = \exp\left[-\frac{s}{L_{H}}\right]\exp\left[\frac{\gamma}{2}(M_{0}^{2}-M^{2}) \right]
\end{equation}

\noindent where $M = \sqrt{\frac{\mu v^{2}}{2\gamma k_{B}\bar{T}}}$, and $L_{H} = -\frac{2k_{B}\bar{T}}{\mu g_{||}}$. The minus sign in $L_{H}$ is due to $g_{||}$ being negative. $M_{0}$ denotes Mach number at base $s$ = 0. 

We now convert the above expression into a form which can be used to place bounds on pressure at base of corona in terms of quantities which can be computed within the domain of 0D description of loops, viz. average pressure and Mach number at coronal base. 
We consider two functions $X_{1}$ and $X_{2}$ of time and loop length such that for a particular loop of length L, the minimum and maximum values of $\exp\left[-\frac{\gamma}{2}(M_{0}^{2}-M^{2}) \right]$ at a time are $\exp\left[X_{1}M_{0}^{2}\right]$ and $\exp\left[X_{2}M_{0}^{2}\right]$ respectively. Hence, averaging equation~\ref{eq:steadyflow2} over the coronal part of the loop, we obtain

\begin{equation}\label{eee}
\left[\frac{P_{0}}{\bar{P}}\right]_{hse}\exp\left[X_{1}M_{0}^{2}\right] \leq \left[\frac{P_{0}}{\bar{P}}\right] \leq \left[\frac{P_{0}}{\bar{P}}\right]_{hse}\exp\left[X_{2} M_{0}^{2}\right]  
\end{equation}

where $\left[\frac{P_{0}}{\bar{P}}\right]_{hse}$ is the ratio of base and average pressure, if system was in hydrostatic equilibrium and isothermal with a temperature $\bar{T}$, while $\left[\frac{P_{0}}{\bar{P}}\right]$ is actual value. The expressions in equation~\ref{eee} is consistent with the fact that in the regime where Mach numbers approach 0, we will recover the hydrostatic expression. 

\begin{figure}[ht]
\centering 
\includegraphics[width=0.5\linewidth]{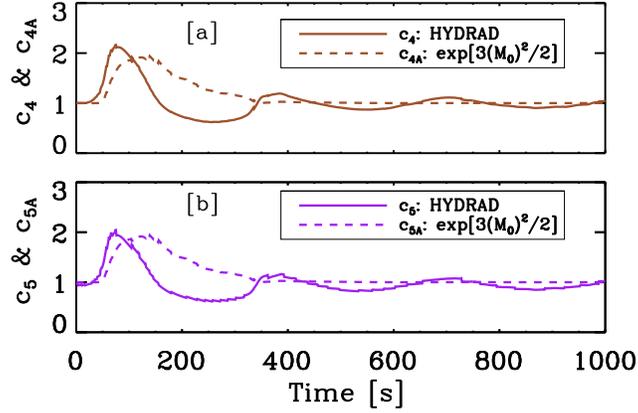} 
\caption{As in panels [a] and [b] of Figure~\ref{fig:figcase01}, but comparing the \hydrad\ pressure and electron number density coefficients $c_{4}$ and $c_{5}$ (solid curves) with approximations made as $[\exp\left(\frac{3}{2}M_{0}^{2}\right)]$ (dashed curves; see equation~\ref{eq:ansatz1}) where the Mach number at the base of the corona, $M_0$, is computed via \hydrad.  Note that the modifications offer a qualitative improvement in the approximations.}

\label{fig:figcase02} 
\end{figure}

Without any loss of generality, we can
express the ratio $\left[\frac{P_{0}}{\bar{P}}\right]$ and  $\left[\frac{n_{0}}{\bar{n}}\right]$

\begin{equation}\label{eq:ansatz}
\left[\frac{P_{0}}{\bar{P}}\right] =  \left[\frac{P_{0}}{\bar{P}}\right]_{hse}\exp[\phi(t,L) M_{0}^{2}]\\
\hspace{0.3cm} \& \hspace{0.3cm} 
\left[\frac{n_{0}}{\bar{n}}\right] =  \left[\frac{n_{0}}{\bar{n}}\right]_{hse}\exp[\phi(t,L) M_{0}^{2}]
\end{equation}

\noindent where $\phi$ is a function of time and loop length. Moreover, for a particular loop at a time t, it is bounded by the relation $X_{1} \leq \phi \leq X_{2}$. 
Even though we arrived at  equation~\ref{eq:ansatz} assuming  steady flow in an
isothermal loop, 
one can always invert it to express $\phi(t,L)$ at each time step in terms of $\left[\frac{P_{0}}{\bar{P}}\right]$, and $\left[\frac{P_{0}}{\bar{P}}\right]_{hse}$, for the general case of non-steady flows in multi-thermal loops. However, it is infeasible 
to derive an exact expression for $\phi(t,L)$ 
as 
it would require a field aligned simulation in the first place to be computed.
Therefore, we approximate it as a constant in time, independent of loop length.  We compared the results between \hydrad\ and \ebtel\ for different values of $\phi$ while also comparing $c_{4}$ and $c_{5}$ with the new code. We compute again the approximations $c_{4A}$ and $c_{5A}$ to coefficients $c_{4}$ and $c_{5}$ using the expressions 
\begin{equation}\label{eq:ansatz1}
c_{4A}  =  \exp[\frac{3}{2}M_{0}^{2}]\\
\hspace{0.3cm} \& \hspace{0.3cm} 
c_{5A} =  \exp[\frac{3}{2}M_{0}^{2}]
\end{equation}
We find that the calculations are not sensitive to the precise value adopted
and choose $\phi=\frac{3}{2}$.  See Figure~\ref{fig:figcase02} for a comparison showing the effect of adopting this nominal value in {\hydrad}.

\begin{figure}[ht]
\centering 
\includegraphics[width=\linewidth]{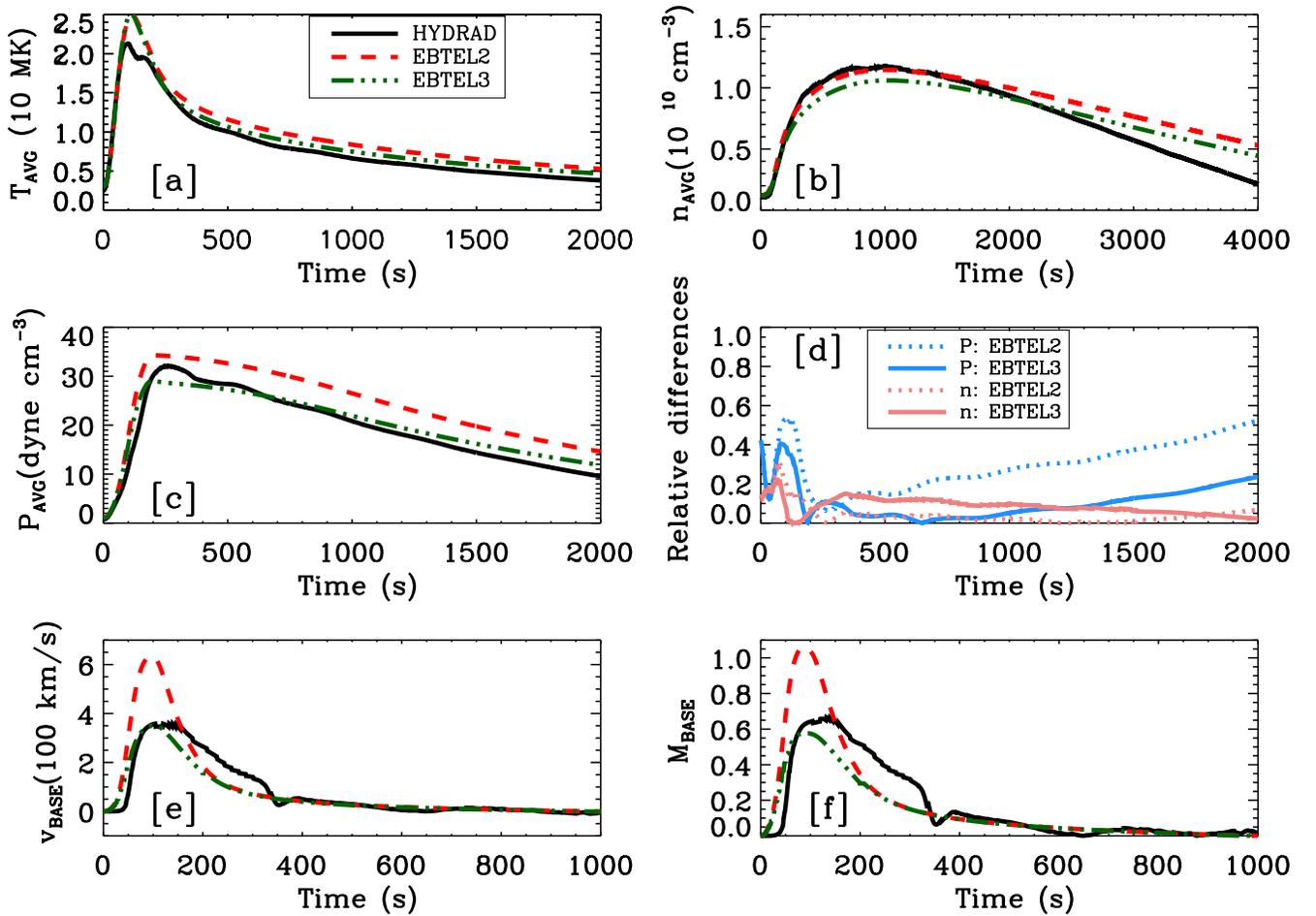} 
\caption{As in  Figure~\ref{fig:figcase0} but with over-plotted green curves obtained with \ebtelll.}\label{fig:figcase1} 
\end{figure}

We plot $c_{4A}$ and $c_{5A}$ in panels [a] and [b] of Figure~\ref{fig:figcase02} using brown dashed ($c_{4A}$) and purple dashed ($c_{5A}$) lines. For comparison we have also plotted $c_{4}$ and $c_{5}$ computed by {\hydrad} using solid lines. The plots reveal that the new approximations on $c_{4}$ and $c_{5}$ resulting into $c_{4A}$ and $c_{5A}$, respectively, shows the most prominent peak observed during the impulsive phase, albeit at slightly later stage. Furthermore, other peaks occurring in $c_{4}$ and $c_{5}$ during the later phase of the evolution are also close to the values of $c_{4A}$ and $c_{5A}$ in the limit $M_{0} \to 0$. However, the prominent dips observed in $c_{4}$ and $c_{5}$ between 150 and 350 sec are not captured by $c_{4A}$ and $c_{5A}$. We have analyzed these shortcomings and found that the errors due to these are rather small because of decent match between profiles of $\bar{P}, \bar{n}, \bar{T}$ and $v_{0}$, computed from field aligned and 0D simulations for different cases considered (see \S~\ref{sec:results} and Figure~\ref{fig:figcase1}).

We stress that the resultant expressions in equation~\ref{eq:ansatz1} are crude, because approximating $\phi(t,L)$ in equation~\ref{eq:ansatz} with a constant value of $\frac{3}{2}$ does not capture its complicated variation with time and loop length. However, it  achieves the twin goals of capturing the most prominent feature (the first peak) of $c_{4}$ and $c_{5}$ computed from {\hydrad} and improving results of 0D simulations over wide range of parameter space of solar coronal loops as long as the flows in {\hydrad} are subsonic.

Finally, using the above described approximations, we simulate the plasma dynamics of the monolithic loop that is discussed in \S\ref{sec:includeke} and plot the obtained results in Figure~\ref{fig:figcase1}.  We denote the results obtained using the above-described approximations, including the addition of KE, with {\ebtelll}. For comparison we have plotted the results obtained from {\hydrad} (black-solid) as well as {\ebtell} (red-dashed). The results show that temperature (see panel [a]) produced by \ebtell~and \ebtelll~are still almost overlapping. However both density and pressure are lower in \ebtelll~than \ebtell.  The density derived using {\ebtell}, however, is better matched with {\hydrad} than that derived with {\ebtelll} (see panel [b]). This discrepancy is partially due to the errors in $c_{4A}$ and $c_{5A}$. However, pressure is better reproduced using {\ebtelll} (see panel [c]).  We plot the velocities and Mach numbers measured at the base of the corona in panels [e] and [f], respectively, that clearly show that {\ebtelll} performs better than {\ebtell}. Though {\ebtell} predicts Mach numbers exceeding unity, {\ebtelll} manages to bring it down to values predicted by \hydrad. We plot the relative errors in average pressure and electron number density in panel [d]. We find that the deterioration in density due to our approximation over {\ebtell} is less than the improvement in pressure.

\subsection{\textbf{Density computed by {\ebtell} and {\ebtelll}}}\label{subsec:compebtel2and3}

The question remains, however, as to why {\ebtell} does a better job than {\ebtelll} in predicting the average density when compared to {\hydrad}, even when the predicted velocity at the coronal base shows substantial discrepancy. This may be explained as follows. During the impulsive phase, conductive losses dominate and radiative losses are negligible \citep[see also][]{rajhans2021hydrodynamics,srividya2018, cargill1995}. Therefore, the sum of enthalpy and kinetic energy flux across coronal base $\left(\frac{\gamma}{\gamma-1}P_{0}v_{0} + \frac{1}{2}\mu n_{0}v_{0}^{2}\right)$ is approximately equal to the conduction flux 
($F_{0}$) across it. This approximate equality can be used in addition to equation~\ref{eq:eos} to write the mass flux across coronal base as

\begin{equation}\label{eq:explanation}
\mu n_{0}v_{0} = -\frac{F_{0}}{\frac{2\gamma k_{B}T_{0}}{\mu(\gamma-1)}+\frac{v_{0}^{2}}{2}}
\end{equation}

We now look at the quantities involved in this equation namely temperature, conduction flux, and velocity. The temperature estimated by {\ebtell} as well as {\ebtelll} are higher than those obtained using {\hydrad}, which is due to conduction flux being under-estimated in {\ebtell} and {\ebtelll}.
Hence a lower magnitude of $F_{0}$ in {\ebtell} and {\ebtelll} than {\hydrad} will tend to make magnitude of $\mu n_{0}v_{0}$ lower in {\ebtell} and {\ebtelll} than {\hydrad}. Additionally an overestimated temperature $T_{0}$ in {\ebtell} and {\ebtelll} will tend to make magnitude of $\mu n_{0}v_{0}$ lower than those obtained from {\hydrad}.  However the neglect of kinetic energy  \revtwo{($\propto \frac{v_{0}^{2}}{2}$)} in {\ebtell} will tend to make the magnitude of mass flux higher than {\hydrad}. This is not the case with {\ebtelll}. This last effect is by far the most important contributor to the discrepancy between mass flux in {\ebtell} and {\hydrad}. The mass flux hence obtained from {\ebtelll} agrees better with {\hydrad}.  

However, neither {\ebtell} nor {\ebtelll} take into account the dynamic change in
the  length of the coronal part of the loop. Both assume that the whole loop is in the corona. To compare the {\ebtell} or {\ebtelll} results, the coronal portion of loops in {\hydrad} is defined dynamically by measuring the length of the part of the loop with temperature $\ge 0.6T_{a}$. Due to this definition,  the coronal portion of the loop identified by {\hydrad} could be significantly smaller than the full length of the loop considered by {\ebtell} and {\ebtelll} in the impulsive phase. This is consistent with the requirement of larger temperature gradients to transfer larger conduction flux to the transition region. The error caused by using larger length of coronal part in equation~\ref{eq:zerodmass} tends to compensate the higher mass flux in {\ebtell}. As a result, $\bar{n}$ computed by {\ebtell} and {\hydrad} match well. The improvement in velocity in {\ebtelll} brings the mass flux computed by {\ebtelll} closer to that by {\hydrad}, but the error caused by taking fixed $L$ in equation~\ref{eq:zerodmass} is not canceled sufficiently. This leads to worsening of densities in {\ebtelll} than {\ebtell}. However, we emphasize that the worsening in densities is significantly smaller than the improvement in pressure and velocity, in terms of agreement with results from {\hydrad}.

\section{Results}\label{sec:results}

Having tested the approximations for a nominal example case, we next carry out a systematic verification covering a useful range of the parameter space of solar coronal loops. All the heating functions have a symmetric triangular profile. The results are compared with those from {\ebtell} and {\hydrad}.  Table~\ref{table:test} provides the details of the test cases chosen for simulations. For the sake of completeness we have also provided in the same table, input parameters for case 1 which was discussed in \S\ref{sec:includeke} and \S\ref{sec:densratio}. Note that the test cases are chosen such that they cover a wide range of loop lengths, heating functions and the maximum velocities reached in {\hydrad} at coronal base  cover the regimes of subsonic (cases 1-4), trans-sonic case (5-6) and supersonic (case 7) flows. The maximum Mach numbers achieved at base ($M_{0}$) in \hydrad~simulations are also listed in table~\ref{table:test}. We discuss the results from all the test cases one by one below.

\begin{table}[h!]
\centering
\caption{Simulation parameters such as peak heating rate (2nd column), duration of heating (3rd column), loop half length  (4th column), initial electron number density (5th column), initial temperature (6th column), maximum Mach numbers at coronal base computed by {\hydrad} (7th column), {\ebtell} (8th column) and {\ebtelll} (9th column) for various test cases.}\label{table:test}
\begin{tabular}{ |c |c |c |c |c |c |c |c |c |c |c| }
\hline
\hline
Index & Peak heating rate  &  Duration of heating  & Half Length  &  Initial $\bar{n}$  &  Initial $\bar{T}$  & (M$_{0}$)$_{max}$     & (M$_{0}$)$_{max}$   & (M$_{0}$)$_{max}$\\ 
 &  [ergs~cm$^{-3}$~s$^{-1}$]  &  [s]  & [10$^{8}$ cm] & [$10^{8}$~cm$^{-3}$] & [$10^{6}$~K]  &   {\hydrad}        & {\ebtell}   & {\ebtelll}\\
\hline
1  &  0.5   & 200.0 & 65.0 & 11.05 & 2.51 & 0.67 & 1.10 & 0.58 \\
2  &  1.5$\times 10^{-3}$  & 500.0 & 75.0 & 0.62  & 0.85  & 0.57 & 0.88 & 0.50 \\
3  &  1.0$\times 10^{-2}$  & 200.0 & 25.0 & 2.46  & 0.73    & 0.49 & 0.61 & 0.42 \\
4  &  2.0   & 200.0 & 25.0 & 22.32 & 2.06   & 0.55 & 0.81 & 0.51 \\
5  &  1.0$\times 10^{-2}$  & 200.0 & 50.0 & 0.84  & 0.71  & 0.75 & 1.32 & 0.63 \\
6  &  1.5$\times 10^{-2}$  & 200.0 & 75.0 & 0.80  & 0.92 & 0.94 & 1.46 & 0.65 \\
7  &  1.0$\times 10^{-3}$  & 200.0 & 60.0 & 0.13  & 0.42 & 1.15 & 1.51 &0.65 \\
\hline 
\end{tabular}
\end{table}

\subsection{Case 2}\label{subsec:case2}

We simulate the plasma dynamics in a loop of half length of 75~Mm with background heating adjusted such that the initial electron number density and temperature are $0.62\times10^{8}$ cm$^{-3}$ and $0.85$ MK, respectively. We provide a symmetric triangular heating profile lasting for 500~s, with the maximum heating rate being 0.0015~ergs~cm$^{-3}$~s$^{-1}$ at $t=250$ s.  

\begin{figure}[ht]
\centering 
\includegraphics[width=\linewidth]{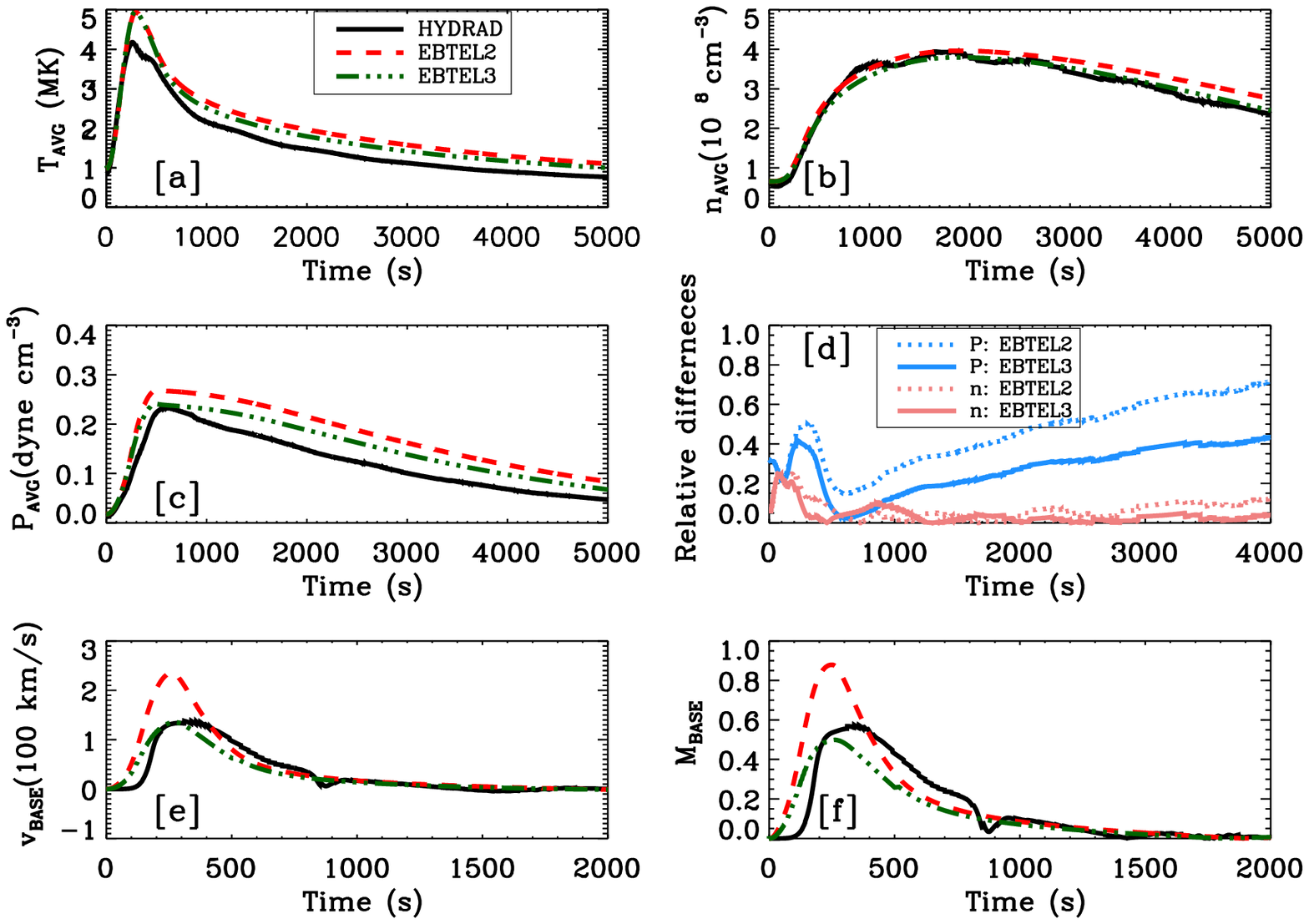} 
\caption{As in 
Figure~\ref{fig:figcase1} but for Case 2: A 75 Mm loop receiving 2.8125$\times10^{9}$ ergs~cm$^{-2}$ in 500 s.}\label{fig:figcase2} 
\end{figure}

The results are plotted in Figure~\ref{fig:figcase2}. As for the first case described above, there is little difference in the temperature profile (panel a), but the electron number density (panel b) is underestimated using {\ebtelll} in comparison with that from {\ebtell} and the match with {\hydrad} deteriorates. Similar to the underestimation of electron number density, pressure is also underestimated with {\ebtelll} and shows better correspondence with the results from {\hydrad}. The relative errors in electron number density and pressure obtained by {\ebtelll} and {\ebtell} are also plotted in panel [d]. Similar to the case 1, we note that the deterioration in the electron number density due to approximation in {\ebtelll} is smaller than the improvements seen in pressure. 

The velocity and the Mach number estimated at the base of the corona, both are lower in simulations using {\ebtelll} in comparison to {\ebtell} and are in excellent agreement with those obtained with {\hydrad}. This is a sub-sonic case, where both {\ebtell} and {\hydrad} produce Mach numbers at the base lower than 1.

\subsection{Case 3}\label{subsec:case3} 

\begin{figure}[ht]
\centering 
\includegraphics[width=\linewidth]{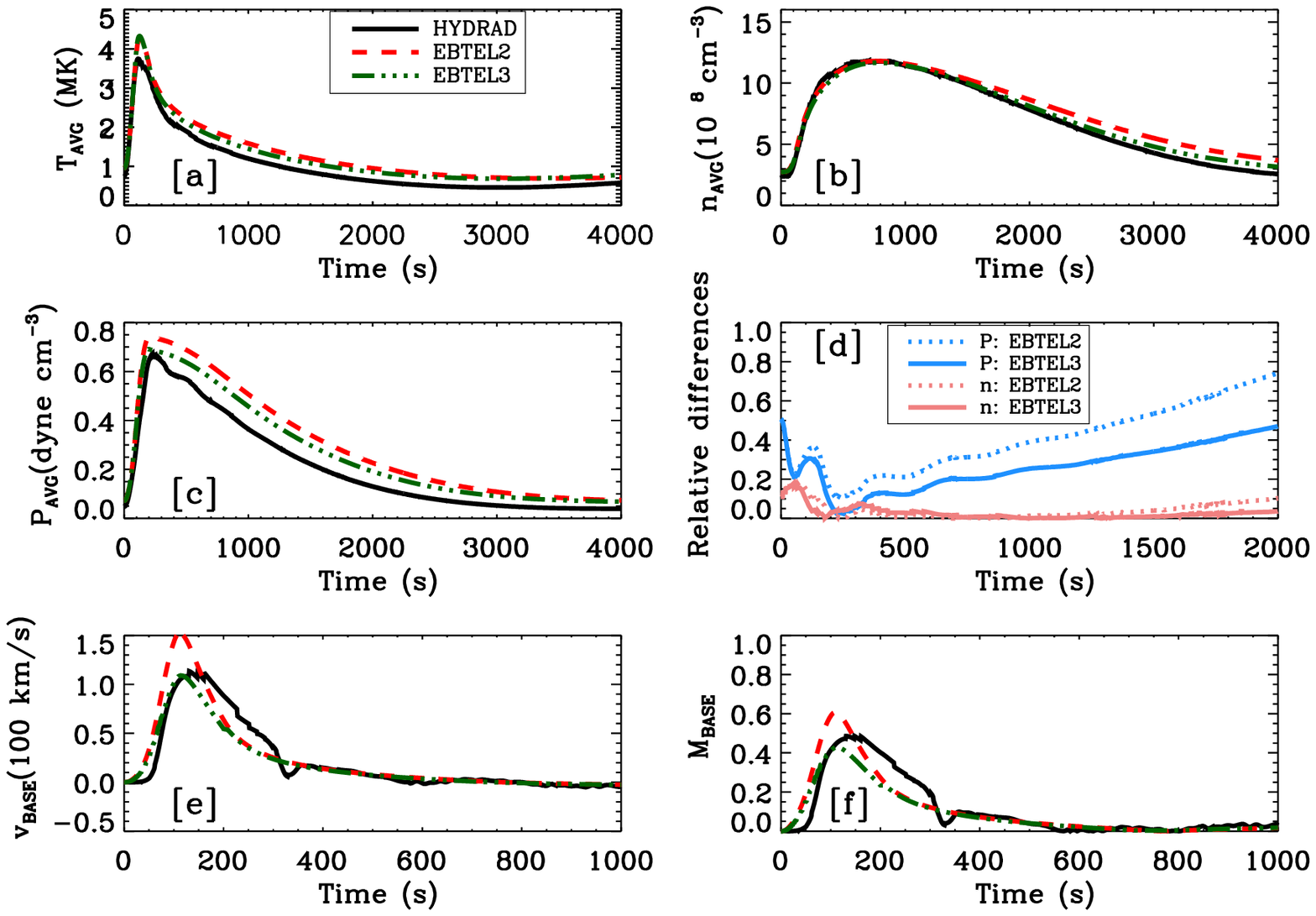} 
\caption{As in  figure~\ref{fig:figcase1} but for  Case 3: A 25 Mm loop receiving 2.5$\times10^{9}$ ergs~cm$^{-2}$ in 200 s.}\label{fig:figcase3}
\end{figure}

The third case simulates the plasma dynamics in a loop of half length of 25 Mm. The background heating is adjusted such that the initial electron number density and temperature are $2.46\times10^{8}$ cm$^{-3}$ and $0.73$ MK respectively. We provide a symmetric triangular heating profile lasting for 200 s, with the maximum heating rate being 0.01 ergs~cm$^{-3}$~s$^{-1}$ at $t=100$ s. We plot the results in figure~\ref{fig:figcase3}. As before, pressure, velocity and Mach number show significant improvements compared with {\ebtell}.

\subsection{Case 4}\label{subsec:case4} 

\begin{figure}[ht]
\centering 
\includegraphics[width=\linewidth]{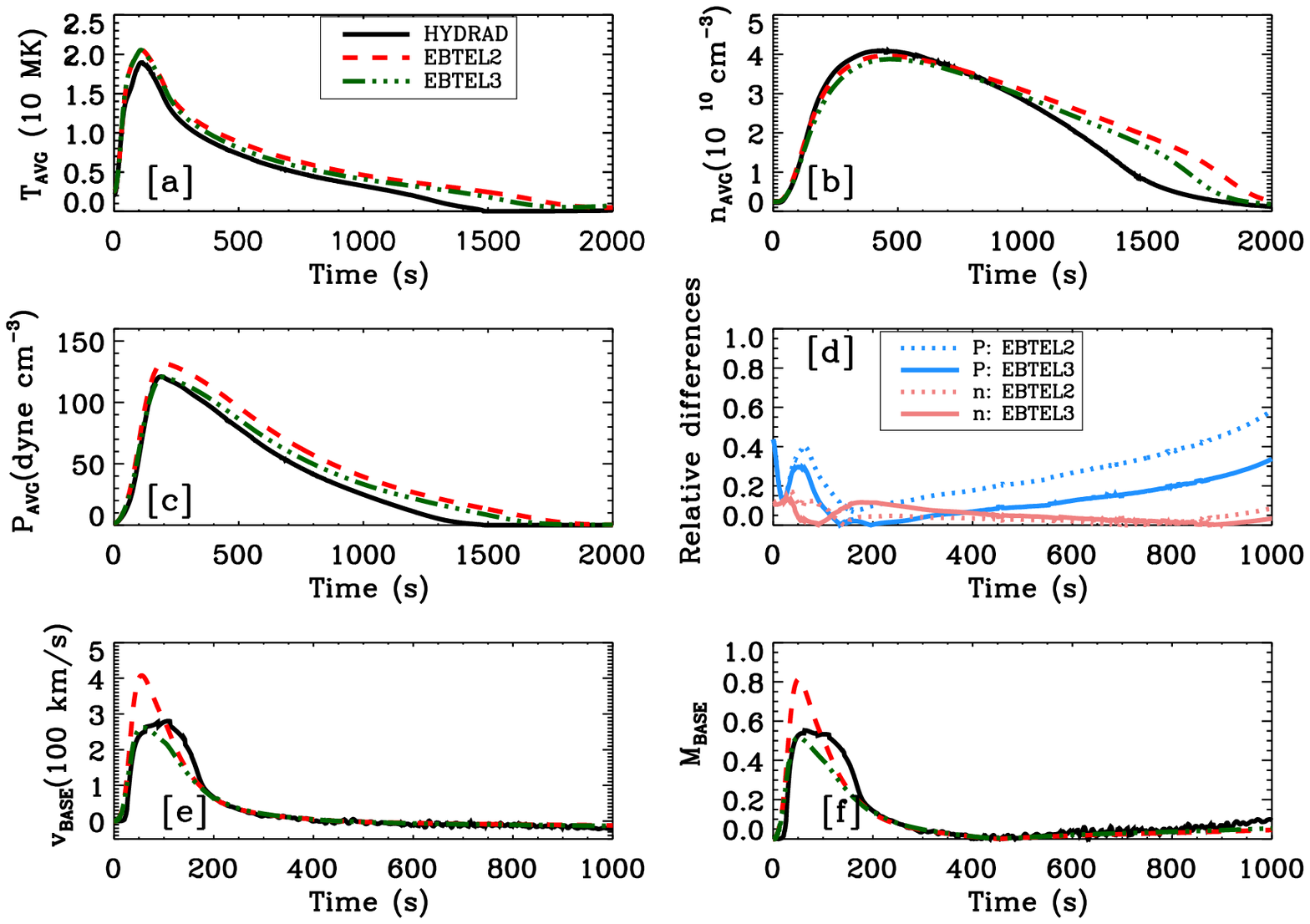} 
\caption{As in  Figure~\ref{fig:figcase1} but for Case 4: A 25 Mm loop receiving 5.0$\times10^{11}$ ergs~cm$^{-2}$ in 200 s.}\label{fig:figcase4} 
\end{figure}

Here we take a loop of half length of 25 Mm with background heating adjusted such that the initial electron number density and temperature are $22.32\times10^{8}$ cm$^{-3}$ and 2.06 MK respectively. We provide a symmetric triangular heating profile lasting for 200 s, with the maximum heating rate being 2.0 ergs~cm$^{-3}$~s$^{-1}$ at $t=100$ s. The results are shown in Figure~\ref{fig:figcase4}. The temperature, electron number density and, pressure show similar evolution as in the other examples described above.  The velocity and Mach number plots shown in panel [e] and [f] shows remarkable improvements in {\ebtelll} over {\ebtell} when compared with {\hydrad}.

\subsection{Case 5}\label{subsec:case5}

\begin{figure}[ht]
\centering 
\includegraphics[width=\linewidth]{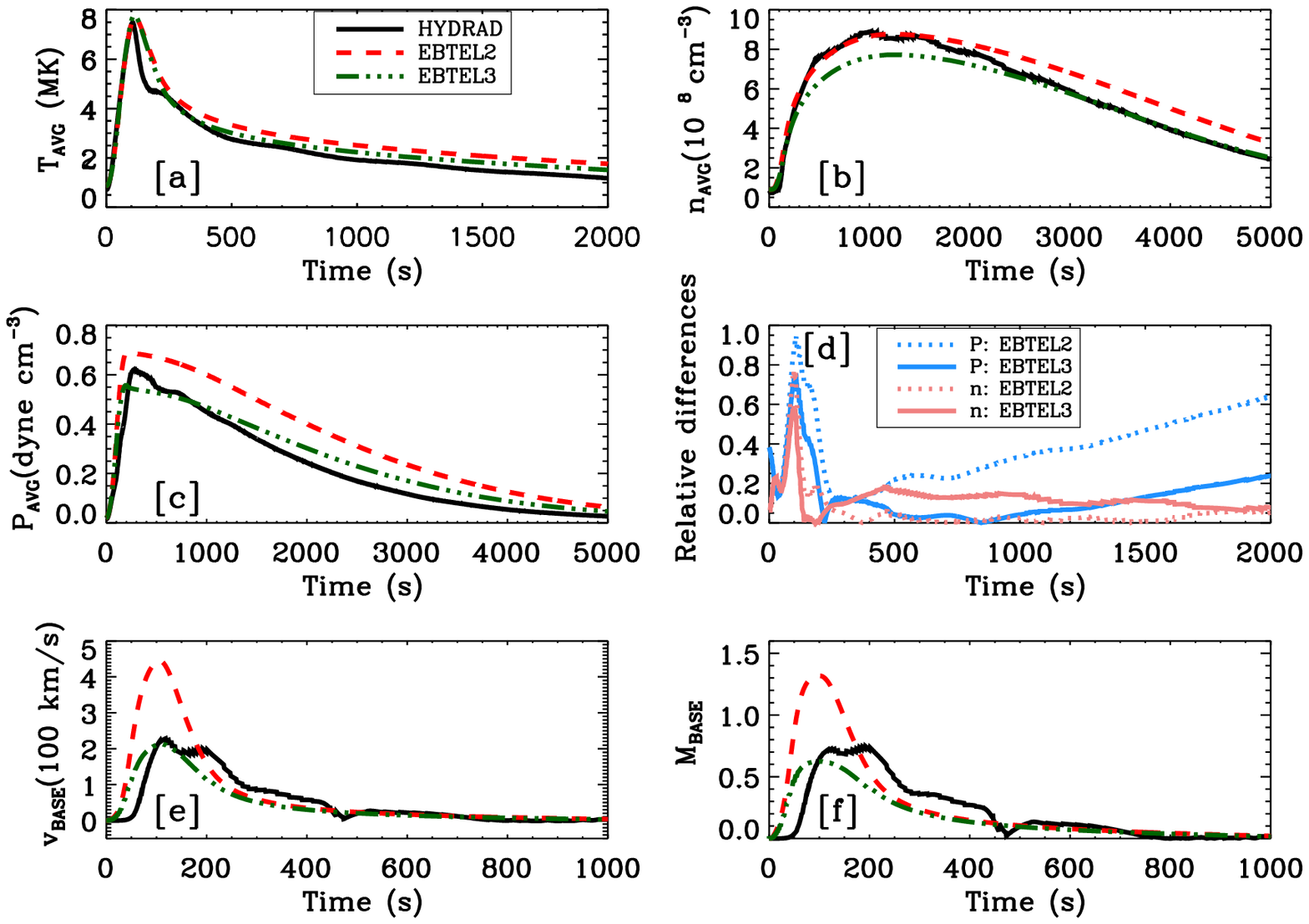} 
\caption{As in Figure~\ref{fig:figcase1} but for Case 5: A 50 $Mm$ loop receiving 5.0$\times10^{9}$ $ergs~cm^{-2}$ in 200 $s$. }\label{fig:figcase5} 
\end{figure}

Here we consider a loop of half length of 50 Mm with initial electron number density and temperature are $0.84\times10^{8}$~cm$^{-3}$ and 0.71~MK. The heating profile lasting for 200 s, with the maximum heating rate being 0.01 ergs~cm$^{-3}$~s$^{-1}$ at $t=100$ s. We plot the results in Figure~\ref{fig:figcase5}. Panel [a] shows the temperature produced by {\ebtell} and {\ebtelll} are not overestimated but in close agreement with {\hydrad}. This is because in this case, even though velocities at base remain subsonic through out, velocities produced by {\hydrad} at intermediate positions in the loop reach supersonic velocities, which leads to shocks and hence dissipation of kinetic energy. This adds to temperature produced by {\hydrad}. Since 0D description cannot incorporate the physics of shocks, there is no increase in temperature due to dissipation of kinetic energy. Panels [b]-[d] show similar trends in electron number density and pressure as in previous cases. Panel [e] and [f] show the velocity and Mach number at coronal base respectively. The maximum Mach numbers produced at base of corona in \hydrad~simulations is 0.75 and those produced by {\ebtelll} are $\approx$ 0.63. Mach number produced in \ebtell~are supersonic $\approx$ 1.4. The match between Mach numbers computed using {\hydrad} and {\ebtelll} is worse than previous cases.

\subsection{Case 6}\label{subsec:case6}

\begin{figure}[ht]
\centering 
\includegraphics[width=\linewidth]{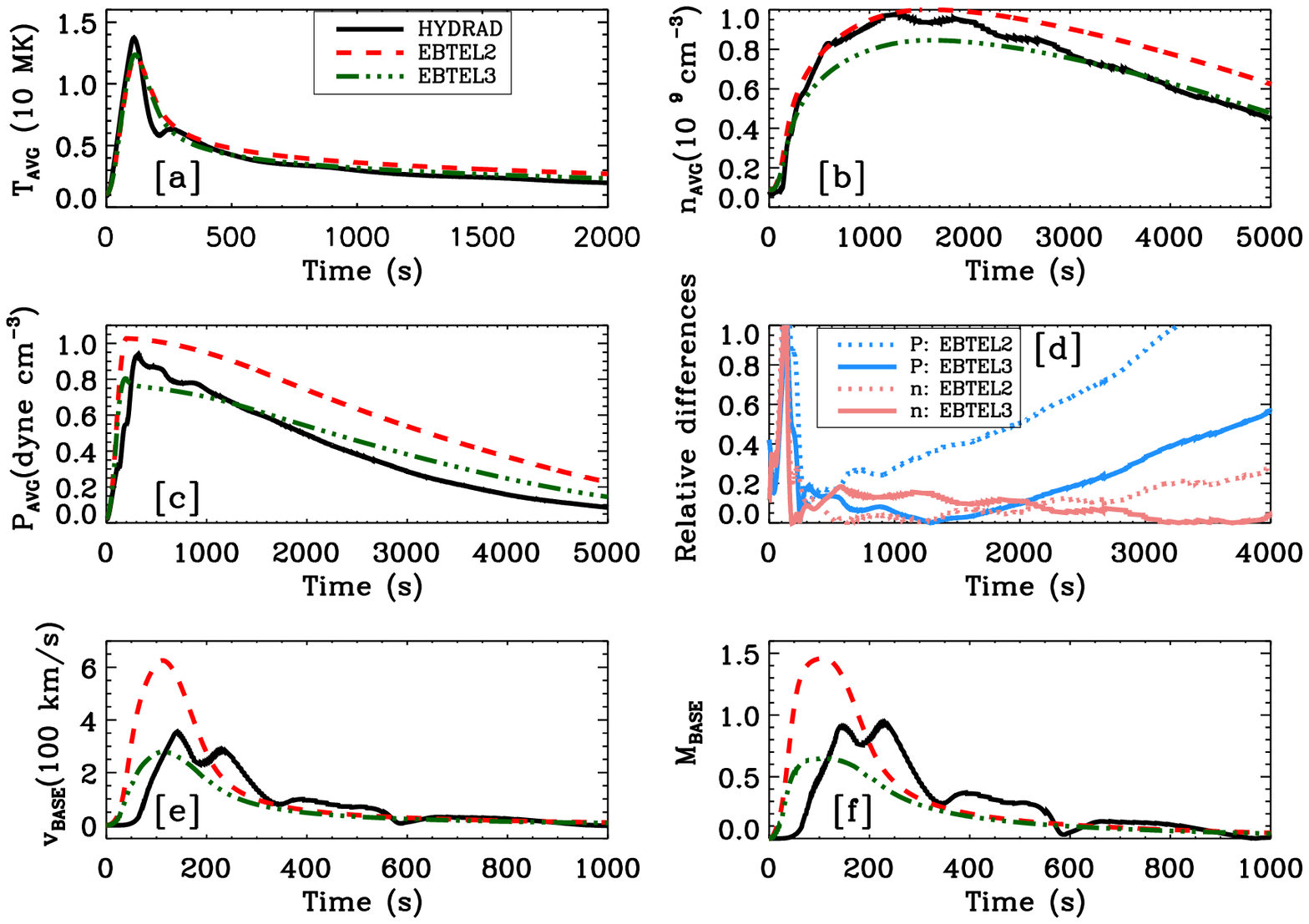} 
\caption{As in Figure~\ref{fig:figcase1} but for Case 6: A 75 Mm loop receiving 1.125$\times10^{10}$ ergs~cm$^{-2}$ in 200 s.}\label{fig:figcase6}
\end{figure}

The sixth case is of a loop of half length of 75 Mm with initial electron number density and temperature are $0.80\times10^{8}$~cm$^{-3}$ and 0.92~MK. The heating profile lasting for 200 s, with the maximum heating rate being 0.015 ergs~cm$^{-3}$~s$^{-1}$ at $t=100$ s. The results are shown in Figure~\ref{fig:figcase6}. {\hydrad} in this case predicts higher temperature than {\ebtel}, due to dissipation of kinetic energy by shocks in {\hydrad} (see panel [a]). We observe the same pattern in the 
time evolution of electron number density and pressure (see panels [b]-[d]). Panel [e] and [f] show the velocity and Mach number at coronal base of loop. In this case, {\hydrad} predicts maximum Mach number of 0.94 while {\ebtell} and {\ebtelll} predicts maximum Mach numbers of 1.5 and 0.65, respectively. Though the Mach numbers produced by {\ebtelll} are unreliable because of being significantly lower than {\hydrad}, the agreement between velocities is better.

Interestingly the agreement between velocity predicted by {\ebtelll} and {\hydrad} is worse in this case than the previous cases where maximum Mach number at coronal base was well below 1. This worsening of velocity computed by {\ebtelll} can be understood as follows. Shocks are formed from the abrupt slowing of a supersonic evaporative upflow. The discontinuous decrease in velocity with height is accompanied by a discontinuous increase in the density.  Clearly, $\frac{n_{0}}{\bar{n}}$ must be higher without the shock than with it. Consequently equation~\ref{eq:ansatz1} gives higher values of $n_{0}$ in {\ebtelll}. The over estimation of $n_{0}$ in equation~\ref{eq:0denergytr_newfinal} leads to a lower value of $v_{0}$.

\subsection{Case 7}\label{subsec:case7}

The seventh case is of a loop of half length of 60 Mm with background adjusted such that the initial electron number density and temperature are $0.13\times10^{8}$~cm$^{-3}$ and 0.42~MK. It receives a symmetric triangular heating profile lasting for 200 s, with the maximum heating rate being 0.001 ergs~cm$^{-3}$~s$^{-1}$ at $t=100$ s. The results are plotted in Figure~\ref{fig:figcase6}.

\begin{figure}[ht]
\centering 
\includegraphics[width=\linewidth]{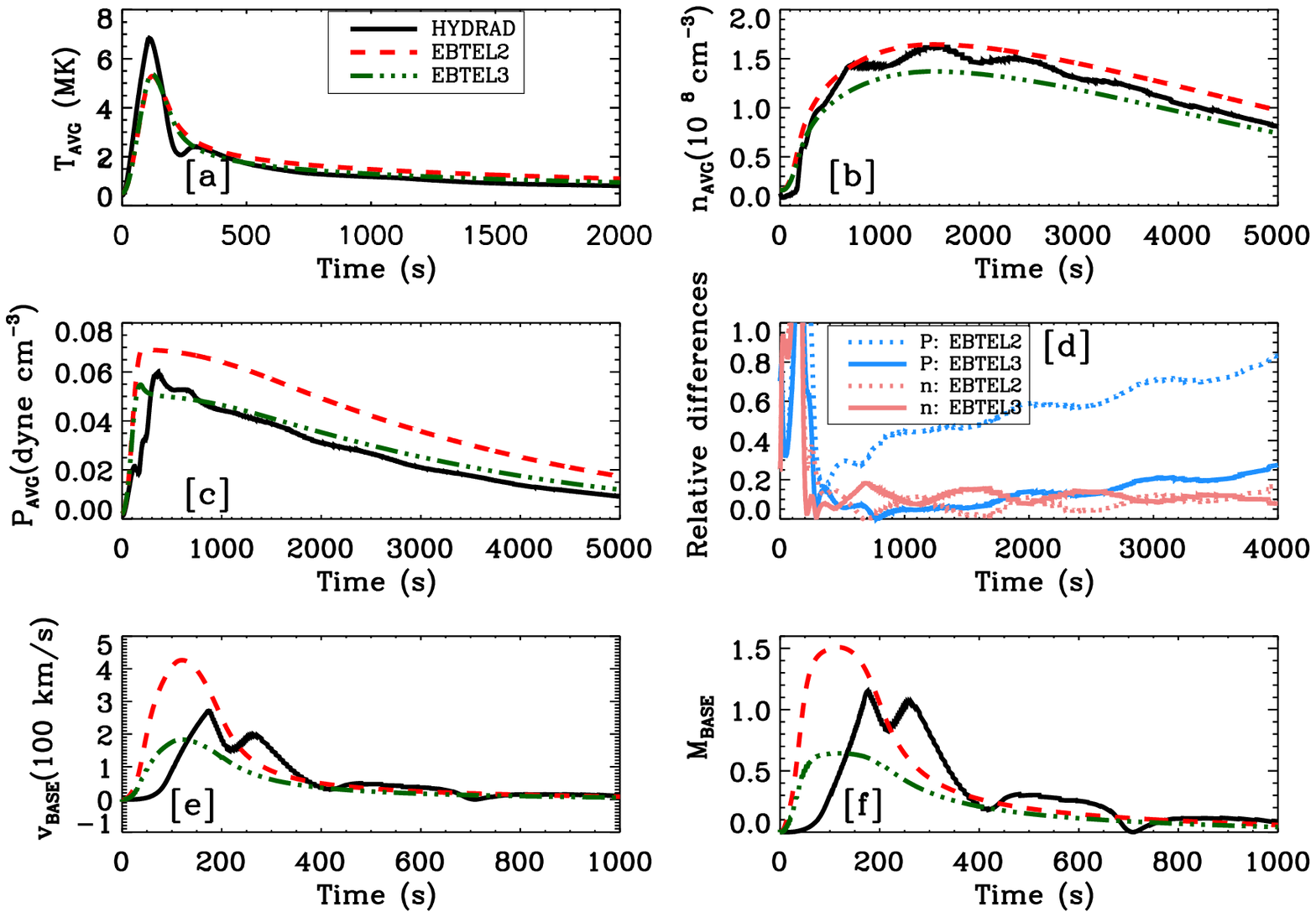} 
\caption{As in Figure~\ref{fig:figcase1} but for Case 7: A 60 Mm loop receiving 6.0$\times10^{8}$ ergs~cm$^{-2}$ in 200 s. }\label{fig:figcase7} 
\end{figure}

In this case, maximum Mach numbers reached by by \hydrad~exceed unity (1.15) at the base. The maximum Mach numbers produced by \ebtell~and \ebtelll~are 1.51 and 0.65 respectively and both are not reliable because 0D simulations cannot handle shocks, which are expected to occur in reality. Despite this, we see that temperature (panel [a]), density (panel [b]) , and pressure (panel [c]) computed from 0D simulations being reliable. As discussed in \S~\ref{subsec:case6}, {\hydrad} computes higher temperatures than {\ebtell} and {\ebtelll} (see panel [a] of Figure~\ref{fig:figcase7}). Panel [e] shows that {\ebtelll} under estimates velocities at coronal base. Even though the peak velocities computed by {\hydrad} match better with {\ebtelll} than {\ebtell}, the Mach numbers computed by {\ebtelll} are subsonic and hence cannot be trusted.

\subsection{Prediction of onset of shocks}\label{subsec:predict}

From the simulations performed with {\ebtell}, {\ebtelll}, and {\hydrad} for various input parameters, we find reasonable agreement between $\bar{P}, \bar{n}$,  and  $\bar{T}$ obtained from {\hydrad} and {\ebtelll}. The agreement for $v_{0}$ and $M_{0}$ is better when maximum $M_{0}$ produced in {\hydrad} is subsonic, but starts worsening when $M_{0}$ starts approaching or exceeding unity. Despite this, we find that that both {\ebtell} and {\ebtelll} 
generate values of $\bar{P}, \bar{n}$,  and  $\bar{T}$ within a factor 2$\times$ that from \hydrad. While {\ebtell} provides better $\bar{n}$, {\ebtelll} provides better $\bar{P}$. Additionally {\ebtelll} does better for velocities. Note however that the match between $M_{0}$ computed from {\hydrad} and {\ebtelll} is not acceptable in the last two cases, where the maximum $M_{0}$ produced in {\hydrad} is $\gtrsim$1, but {\ebtelll} produces subsonic flows (see table.~\ref{table:test}).  This is because such $M_{0}$ in {\hydrad} would lead to shocks, which cannot be modeled by a simple 0D description of loops. Consequently this may lead to the erroneous conclusion that flows in the loop are subsonic. However, we note that even in these
cases, 0D is still sufficient to study the electron number density, temperature and pressure as they are not affected as shown in Figures \ref{fig:figcase6}~and~\ref{fig:figcase7}. This is consistent with qualitative arguments made in \cite{cargill0dmodels}.

In light of above arguments it becomes necessary to be able to use results from 0D simulations for predicting regimes where the maximum $M_{0}$ produced in field aligned simulations is $\gtrsim$1. In such cases detailed field aligned hydrodynamical simulations are better suited to compute 
reliable Mach numbers. 

Based on the {\ebtelll} simulations,  we find a simple criterion that can be employed to predict this regime. A measurement of the ratio of full width at half maximum (FWHM) of the 
Mach number profile predicted by {\ebtelll} and that of the heating functions
may be used as a diagnostic;  when this ratio $\gtrsim$2, flows in \hydrad\ approach or exceed supersonic velocities. Since the heating functions have symmetric triangular profiles, their FWHM is simply half of the total duration. $M_{0}$ profiles, however, have no concrete shape, hence we resort to finding it numerically by implementing the definition of FWHM.
We tabulate these ratios in table~\ref{table:Mach} for each case along with the maximum Mach number predicted by {\hydrad}. A shock should produce a local increase in average temperature in the corona, i.e., a local peak at some time apart from the peak which corresponds to maximum direct heating. The presence of such local peaks in $\bar{T}$, and sonic-supersonic flows at 
the base in {\hydrad} is seen in cases where  the ratio of FWHMs of $M_{0}$ and the 
heating function is $\gtrsim$ 2. In such cases there is a large  departure from $M_{0}$ predicted using {\ebtelll} results. Hence if this ratio $\gtrsim$ 2, the Mach numbers obtained with \ebtelll\ should be treated with caution.

\begin{table}
\centering 
\caption{Ratio of FWHM of profiles of Mach number at base and heating function }\label{table:Mach}
\begin{tabular}{ |c |c |c |c |c |}
\hline
\hline
Case & Maximum value of M$_{0}$ in \hydrad &  FWHM of M$_{0}$ profile in \ebtelll (r$_{1}$) & FWHM of input heating profile (r$_{2}$) & r$_{1}$/r$_{2}$ \\ 
\hline
1 & 0.67 & 165 & 100 & 1.65 \\
2 & 0.57 & 396 & 250 & 1.58 \\
3 & 0.49 & 155 & 100 & 1.55 \\
4 & 0.55 & 120 & 100 & 1.20 \\
5 & 0.75 & 193 & 100 & 1.93 \\
6 & 0.94 & 227 & 100 & 2.27 \\
7 & 1.15 & 259 & 100 & 2.59 \\
\hline 
\end{tabular}
\end{table}

\section{Summary and Discussion}\label{sec:disc}

Understanding the complete plasma dynamics in coronal loops is important for 
understanding the energetics of the corona. Under the assumption of an absence of cross-field conduction, these loops can be modeled reasonably well using field aligned hydrodynamical simulations viz., {\hydrad}. However, if estimates of the 
evolution of loops in more realistic scenarios where thousands of elemental strands are present and multiple heating events are involved, field aligned simulations are computationally expensive. Even for cases of a monolithic loop, 
obtaining quick and approximately accurate estimates of loop evolution over a wide range of parameters, 0D simulations like \ebtel\ provide a useful alternative.

\cite{cargill} and \cite{Klimchuk2008} had assumed subsonic flow 
exists at all stages and hence neglected the kinetic energy term 
in the energy equation. While this assumption holds good during most of the evolution of loop, it fails during the impulsive phase of some of the heating events. In this paper, we have relaxed the assumption of subsonic flows by not neglecting the kinetic energy term in the energy equation. 

In order to solve the equations, we have made the following two assumptions:
 $${\rm (i)}~ \left[\frac{P_{0}}{\bar{P}}\right] = \left[ \frac{P_{0}}{\bar{P}}\right]_{hse}\left[\exp\left(\frac{3}{2}M_{0}^{2}\right)\right] \implies  \left[\frac{n_{0}}{\bar{n}}\right] = \left[ \frac{n_{0}}{\bar{n}}\right]_{hse}\left[\exp\left(\frac{3}{2}M_{0}^{2}\right)\right] $$ 
 $${\rm (ii)}~ \int_{0}^{L}v\frac{\partial P}{\partial s}ds \approx - \frac{1}{2}n_{0}\mu v_{0}^{3}~~~~~~~~~~~~~~~~~~~~~~~~~~~~~~~~~~~~~~~~~~~~~~~~~~~$$ 
The results obtained with respect to the plasma dynamics in different kinds of loops by {\ebtelll} show significant improvements in average pressure in corona and predicted velocities at coronal base, compared with the results obtained with field aligned simulations using {\hydrad}. 
Though the electron number density produced by \ebtelll~are less accurate than those produced by \ebtell, the discrepancy still remains less than 20\%. The improvement made in pressure estimates by \ebtelll~are larger than the deterioration in electron number density estimates. The main improvement of \ebtelll\ over \ebtell\ is however that of velocities, which match better with \hydrad~results. Additionally  \ebtelll~guarantees $M_{0}$ to remain subsonic if $M_{0}$ produced by \hydrad~remain subsonic.

Furthermore, we have developed a simple heuristic to check whether field aligned simulations produce subsonic flows without performing field aligned simulations. This is useful in deciding if the Mach numbers computed by 0D simulations can be trusted, because 0D simulations are not designed to tackle supersonic 
plasma flows which lead to complicated situations like shocks. For this we look at the ratio of FWHM of profiles of $M_{0}$ and heating profiles. If the ratio is larger than 2, Maximum Mach numbers produced at base of corona are close to or even larger than 1. Nevertheless, even 
in the cases where {\hydrad} predicts trans-sonic and supersonic flows, and the Mach numbers derived by {\ebtell} and {\ebtelll} cannot be trusted, we find the coronal averages ($\bar{T}, \bar{P}$ and $\bar{n}$) calculated by {\ebtelll} to be in good agreement with {\hydrad}.
\acknowledgements
We thank the referee for many comments and suggestions which significantly increased the lucidity of the manuscript. We thank Dr.\ Peter Cargill for his valuable comments and suggestions that improved the manuscript. We also thank Dr.\ Will Barnes for bringing to our attention, a plot showing evolution of various energy terms with time in \cite{Barnes_2016}.  AR acknowledges financial support from University Grants Commission in form of SRF. This work is partly supported by the Max-Planck Partner Group on "Coupling and Dynamics of
the Solar Atmosphere" of MPS at IUCAA. VLK acknowledges support from NASA Contract NAS8-03060 to the Chandra X-ray Center, and the hospitality of IUCAA during several visits. The work of JAK was supported by the Internal Scientist Funding Model (competitive work package program) at Goddard Space Flight Center.

\software{IDL~(\url{https://www.l3harrisgeospatial.com/Software-Technology/IDL}, and \ebtel~\citep{Klimchuk2008,cargill,Barnes_2016}}

\appendix
\section{Analytical solution to cubic Equation}\label{app:cubic}
Here, we
discuss calculation of velocities in {\ebtelke}. Using $P_{0} = 2n_{0}k_{B}T_{0}$, we can re-arrange equation~\ref{eq:0denergytr_newfinal} in the following form
\begin{equation}\label{eq:treqrearranged}
0 =  v_{0}^{3}+\frac{4\gamma k_{B}T_{0}}{\mu(\gamma-1)}v_{0} + \frac{2}{\mu n_{0}}\left(F_{0}+c_{1}\bar{n}^{2}\Lambda(\bar{T})L\right)
\end{equation}

The generic cubic equation given by $v_{0}^{3}+av_{0}^{2}+bv_{0}+c=0$, for real a, b and c can either have three real roots, only one real root, or repeated real roots (i.e. two roots being same). Two quantities f and q are defined as 
\begin{equation}\label{eq:define_fq}
f=b-\frac{a^{2}}{3} =\frac{4\gamma k_{B}T_{0}}{\mu(\gamma-1)} \\
\hspace{0.3cm} \& \hspace{0.3cm} 
q=\frac{2a^{3}}{27}-\frac{ab}{3}+c   = \frac{2}{n_{0}\mu}\left(F_{0}+c_{1}\bar{n}^{2}\Lambda(\bar{T})L\right) 
\end{equation}  
In above equation we have used the fact that coefficient of $v_{0}^{2}$ is 0 i.e. $a=0$.

A discriminant ($\bigtriangleup$) can be defined as
\begin{equation}\label{eq:discriminant}
\bigtriangleup=\frac{q^{2}}{4}+\frac{f^{3}}{27}= \frac{c^{2}}{4}+\frac{b^{3}}{27}
\end{equation}
If $\bigtriangleup$ is positive we have only one real root which is given by
\begin{equation}\label{eq:onedistrealrt}
v_{0} = \left(-\frac{q}{2}+\sqrt{\bigtriangleup}\right)^{\frac{1}{3}}+\left(-\frac{q}{2}-\sqrt{\bigtriangleup}\right)^{\frac{1}{3}}
\end{equation}
For real c, $c^{2}$ hence $q^{2}$ cannot be negative. $f^{3}$ is also positive. Hence we get only one single real root.

However, this method cannot be used directly in {\ebtelll}. Due to the modifications made to the expression of the base pressure and hence base electron number density in section~\ref{sec:densratio} (see equation~\ref{eq:ansatz1}), equation~\ref{eq:treqrearranged} is no longer cubic, and has the form
$$ v_{0}(t)^{3} + b\,v_{0}(t) + c\,\exp({d\,v_{0}(t)^{2}}) = 0 \,.$$

Therefore, to solve the above equation, we first use an adaptive time grid where the time step ${\delta}t$ is set to $0.1$ times the smallest of the conductive, radiative, and sonic timescales at time $t$.
At each time step $t$, we use the velocity at 
the 
previous time step ($v_{0}(t-\delta t$)) to find 
the 
analytical root of the equation cubic in $v_0(t)$, i.e.,
$$v_{0}(t)^3 + b\,v_{0}(t) + c\,\exp(d\,v_{0}(t-\delta t)^{2}) = 0 \,, $$

The values of the physical quantities obtained in consecutive time steps differ by $<$10\% due to the adaptive time stepping. Note that since the loop evolution always begins from a stationary state, i.e., at $t=0$, velocity is zero.

\bibliography{modebtel}
\end{document}